\documentclass[amsmath, prl, twocolumn, superscriptaddress, showpacs,aps, 10pt]{revtex4-1}
\usepackage{graphicx}

\begin{document}     

\title{Quantum dynamics of an atom orbiting around an optical nanofiber}
 
\author{Fam Le Kien}
\email{kien.pham@ati.ac.at}
\altaffiliation{also at Institute of Physics, Vietnamese Academy of Science and Technology, Hanoi, Vietnam.}
\affiliation{Center for Photonic Innovations, 
University of Electro-Communications, Chofu, Tokyo 182-8585, Japan}
\affiliation{Vienna Center for Quantum Science and Technology, TU Wien -- Atominstitut, Stadionallee 2, 1020 Vienna, Austria}

\author{K. Hakuta} 

\affiliation{Center for Photonic Innovations, 
University of Electro-Communications, Chofu, Tokyo 182-8585, Japan}

\author{P. Schneeweiss}

\author{A. Rauschenbeutel} 

\affiliation{Vienna Center for Quantum Science and Technology, TU Wien -- Atominstitut, Stadionallee 2, 1020 Vienna, Austria}

\date{\today}

\begin{abstract}
We propose a novel platform for the investigation of quantum wave packet dynamics, offering a complementary approach to existing theoretical models and experimental systems. It relies on laser-cooled neutral atoms which orbit around an optical nanofiber in an optical potential produced by a red-detuned guided light field. We show that the atomic center-of-mass motion exhibits genuine quantum effects like collapse and revival of the atomic wave packet. As distinctive advantages, our approach features a tunable dispersion relation as well as straightforward readout for the wave packet dynamics and can be implemented using existing quantum optics techniques.
\end{abstract}

\maketitle
Since the earliest days of quantum mechanics, the study of localized, time-dependent solutions to bound-state problems has attracted considerable attention~\cite{Schroedinger26}. Quantum mechanical objects show both particle-like and wave-like behavior and this duality can be readily explored by analyzing the dynamics of wave packets. Such localized states behave like classical objects and follow classical trajectories as long as they do not disperse. However, for nonlinear dispersion relations, the wave packet will spread out and the dynamics cannot be described anymore using classical physics. One of the most prominent witnesses of genuine quantum mechanical dynamics is when the collapse of the wave packet is followed by a revival, i.e., a relocalization of the wave function~\cite{Stroud1986, revival}. This intriguing quantum effect has been experimentally observed in a number of systems, see~\cite{revival} and references therein, and dispersion engineering of wave packets as well as their collapse and revival dynamics constitute an active field of current research \cite{Will10, Wyker12, Keitel}.

Here, we propose a novel experimental platform for the investigation of quantum wave packet dynamics, offering a complementary approach to existing theoretical models and experimental systems. Our approach features a tunable dispersion relation as well as straightforward readout for the wave packet dynamics and can be implemented using existing quantum optics techniques. It relies on laser-cooled neutral atoms which are interfaced with an optical nanofiber as initially proposed in Ref.~\cite{Dowling} and thoroughly analyzed in Ref.~\cite{onecolor}. In this scenario, an atom orbiting around the nanofiber can show quantum mechanical dynamics including an initially classical orbiting motion of the atomic wave packet around the nanofiber, a spread of the atomic wave function (collapse), a partial relocalization of the atom (fractional revival) \cite{Averbukh}, and, finally, a full quantum revival of the original wave packet. In the following, we study this dynamics quantitatively, give analytical expressions for relevant time scales, and derive the functional dependence of the expected signals when absorptively probing the dynamics using a nanofiber-guided, resonant light field.

\begin{figure}
\begin{center}
  \includegraphics[width=5 cm]{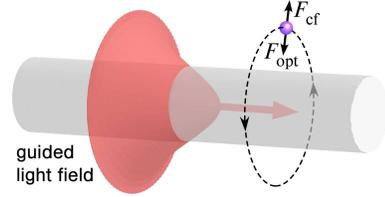}
 \end{center}
\caption{Schematic of trapping an atom in an orbit around an optical nanofiber by the evanescent wave of a quasi-circularly polarized red-detuned guided light field.}
\label{fig1}
\end{figure}

Consider an atom moving in vacuum around a silica nanofiber in a cylindrically symmetric potential $U(r)$, see Fig.~\ref{fig1}. We use cylindrical coordinates $\{r,\varphi,z\}$, with $z$ being the axis of the nanofiber.
Due to the cylindrical symmetry of the system, the angular momentum $L_z=-i\hbar\partial/\partial\varphi$ of the atom with respect to the fiber axis $z$ is conserved. In the eigenstate problem, we have $L_z=\hbar m$, where $\hbar$ is the reduced Planck constant and $m$ is an integer, called the azimuthal quantum number.
Hence, the atom's center-of-mass (COM) motional eigenstates 
can be written as
$\Psi_{\nu mK}=(2\pi)^{-1}R_{\nu m}(r)e^{im\varphi}e^{iK z}$.
Here, $\nu$ is the radial vibrational quantum number,  $K$ is the wave number of the matter wave along the $z$ direction, and $R_{\nu m}(r)$ is the radial part of the wave function. 
We perform the transformation $R_{\nu m}(r)=u_{\nu m}(r)/\sqrt{r}$.
The function $u_{\nu m}(r)$ is determined by the equation
\begin{equation}
\left[-\frac{\hbar^2}{2M}\frac{\partial^2}{\partial r^2}
+U_{\mathrm{eff}}^{(m)}(r)\right]u_{\nu m}(r)=\mathcal{E}_{\nu m} u_{\nu m}(r),
\label{4}
\end{equation}
where $M$ is the atomic mass,
$U_{\mathrm{eff}}^{(m)}(r)=U_{\mathrm{cf}}^{(m)}(r)+U(r)$
is the effective potential, with the centrifugal potential
$U_{\mathrm{cf}}^{(m)}(r)=\hbar^2(m^2-1/4)/(2Mr^2)$, while $\mathcal{E}_{\nu m}$ is the energy eigenvalue for the COM motion of the atom transverse to the fiber. Thus, the  radial motion of the atom orbiting around the nanofiber can be reduced to the motion of a particle in the one-dimensional effective potential  $U_{\mathrm{eff}}^{(m)}(r)$. 

There exist stable bound states for the atom if the effective potential
$U_{\mathrm{eff}}^{(m)}$ has a local minimum outside of the fiber \cite{onecolor}. This may happen if
$U$ is attractive, opposite to the centrifugal potential $U_{\mathrm{cf}}^{(m)}$. 
In order to produce a cylindrically symmetric attractive potential, we send a circularly polarized red-detuned optical field of frequency $\omega$ through the nanofiber. When the nanofiber radius $a$ is small enough, the fiber can support only the fundamental mode HE$_{11}$ \cite{fiber books}, which generates an evanescent-wave guided light field around the nanofiber.

The optical potential for the atom is given by $U_{\mathrm{opt}}=-\alpha(\omega) |\mathbf{E}|^2/4$, where $\alpha(\omega)$ is the real part of the atomic polarizability at the optical frequency $\omega$ and $\mathbf{E}$ is  the positive-frequency component of the electric part of the guided light field. The details of the calculations of the optical potential $U_{\mathrm{opt}}$ for a cesium atom in the vicinity of a nanofiber can be found in Refs. \cite{onecolor,twocolors}. We also take into account the attractive fiber-induced van der Waals potential $U_{\mathrm{vdW}}$ of the atom \cite{onecolor,twocolors}. The combination of the optical potential and the van der Waals potential yields the potential $U=U_{\mathrm{opt}}+U_{\mathrm{vdW}}$. 

\begin{figure}
\begin{center}
  \includegraphics{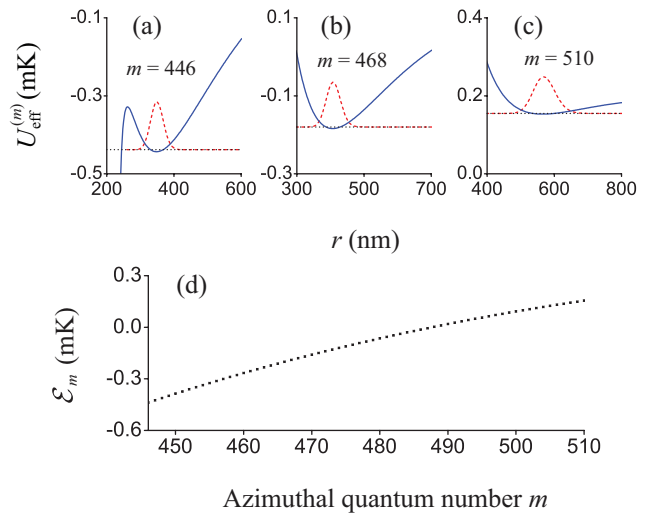}
 \end{center}
\caption{Effective potential $U_{\mathrm{eff}}^{(m)}$ (solid blue lines) and ground-vibrational-state eigenfunction
$u_m$ (dashed red lines) for a ground-state cesium atom orbiting with the azimuthal quantum number $m=446$ (a), 468 (b), and 510 (c) around the nanofiber. (d) Dispersion relation of the ground-vibrational-state energy $\mathcal{E}_m$.
} 
\label{fig2}
\end{figure}

We plot in Figs.~\ref{fig2}(a)--(c) the effective potential $U_{\mathrm{eff}}^{(m)}$ and the ground-vibrational-state eigenfunction $u_m\equiv u_{\nu=1, m}$ for a cesium atom in its electronic ground state orbiting around the nanofiber with the azimuthal quantum number $m$. In the following, the fiber radius is 200 nm and the wavelength and power of the trapping light are 1064~nm and 20~mW, respectively. We note that these parameters are well accessible experimentally \cite{twocolor experiment, Goban12}. Our calculations show that a trapping potential with a minimum outside the fiber can be formed when the azimuthal quantum number $m$ is in the range from 430 to 530. We find that, in this region, the ground-vibrational-state energy $\mathcal{E}_m\equiv\mathcal{E}_{\nu=1,m}$ increases with increasing $m$ and exhibits a small negative curvature, see Fig.~\ref{fig2}(d). We note that we can tune the dispersion relation by varying the power of the guided light field, its wavelength, and/or the fiber radius.

In the following, the motion of the atom along the fiber axis is disregarded because it is independent of the motion in the fiber transverse plane and, consequently, independent of the azimuthal quantum number $m$. Let $|\psi_m\rangle$ be the ground state of the atomic COM motion transverse to the fiber with the angular momentum $\hbar m$. The wave function of this angular momentum state is $\psi_m(\mathbf{r})=(2\pi r)^{-1/2}u_m(r)e^{im\varphi}$ and the corresponding energy is $\mathcal{E}_m$. 
Here, we have introduced the notation $\mathbf{r}=\{r,\varphi\}$. 
Consider a wave packet 
$|\psi\rangle=\sum_m c_m|\psi_m\rangle$,
which is a linear superposition of the angular momentum states $|\psi_m\rangle$ with
the corresponding probability amplitudes $c_m$.
The temporal evolution of this superposition state is given by the wave function
\begin{equation}
\psi(\mathbf{r},t)=\sum_m c_m e^{-i\mathcal{E}_mt/\hbar}\psi_m(\mathbf{r}).
\label{8}
\end{equation}
We assume that the amplitudes of the individual angular momentum states are of the standard Gaussian distribution form,  
$c_m=(2\pi\Delta m^2)^{-1/4}\exp[-(m-m_0)^2/4\Delta m^2]$, with a peak at $m_0$ and a standard deviation $\Delta m$, and are truncated at $m_{\mathrm{min}}<m_0$ and  $m_{\mathrm{max}}>m_0$. In our numerical calculations, we use $m_0=468$, $\Delta m=6$, $m_{\mathrm{min}}=446$, and $m_{\mathrm{max}}=510$. 

Since we have $m_0 \gg \Delta m \gg 1$, we can expand $\mathcal{E}_m$ in a Taylor series around $m_0$ according to
\begin{equation}
\mathcal{E}_m\cong \mathcal{E}_{m_0}+\mathcal{E}_{m_0}' (m-m_0)+\mathcal{E}_{m_0}'' (m-m_0)^2/2~,
\label{Taylor}
\end{equation}
where $\mathcal{E}'_m=d\mathcal{E}_m/dm$ and $\mathcal{E}''_{m}=d^2\mathcal{E}_m/dm^2$. In our case, we have $|\mathcal{E}''_{m_0}|\ll \mathcal{E}'_{m_0}$ which leads to the existence of two different time scales.  Inserting Eq.~(\ref{Taylor}) into Eq.~(\ref{8}) then allows one to identify these relevant time scales of the evolution of the wave packet \cite{revival}. 
The classical period of the rotation of the wave packet in the fiber transverse plane is given by 
$T_{\mathrm{rot}}=2\pi\hbar/\mathcal{E}'_{m_0}$. The characteristic collapse time is defined as the time at which the spread of phase differences between the various oscillatory terms in Eq.~(\ref{8}) is about $\pi$, that is, when the interference is most destructive. This characteristic time is given by 
$T_{\mathrm{coll}}=2\sqrt{\pi}\hbar/(|\mathcal{E}''_{m_0}|\Delta m)$. The revival time for the atomic wave packet is defined as the time at which the phase difference between two neighboring terms in Eq.~(\ref{8}) is $2\pi$, that is, when the interference is most constructive. It is given by $T_{\mathrm{rev}}=4\pi\hbar/|\mathcal{E}''_{m_0}|$. 
For the parameters used in our numerical calculations, we have
$\mathcal{E}'_{m_0}/2\pi\hbar\cong 214$ kHz and $|\mathcal{E}''_{m_0}|/2\pi\hbar\cong 2.52$ kHz, which lead to $T_{\mathrm{rot}}\cong 4.67$ $\mu$s, $T_{\mathrm{coll}}\cong 37.3$ $\mu$s, and $T_{\mathrm{rev}}\cong 794$ $\mu$s.

\begin{figure}
\begin{center}
  \includegraphics{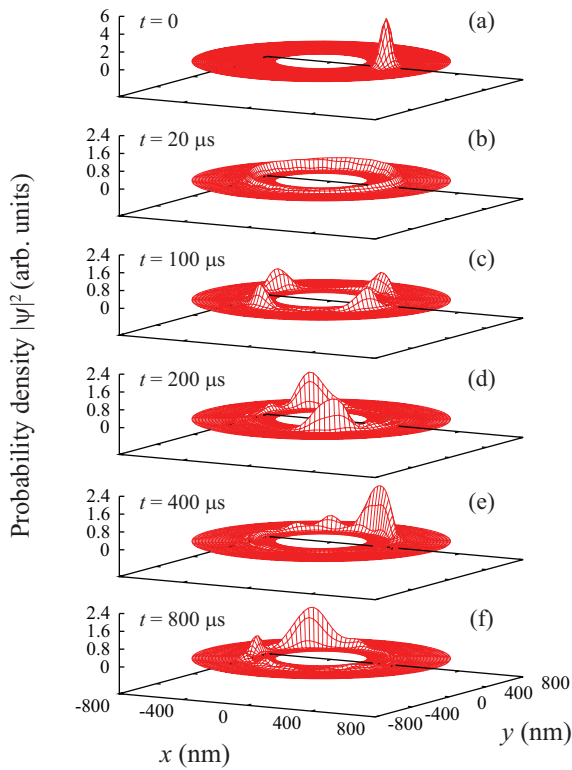}
 \end{center}
\caption{Atomic probability density $|\psi(\mathbf{r},t)|^2$ in the fiber transverse plane
at different evolution times. 
$|\psi(\mathbf{r},t=0)|^2$ is a Gaussian wave packet of angular momentum states with $m_0=468$, $\Delta m=6$,  $m_{\mathrm{min}}=446$, and  $m_{\mathrm{max}}=510$.
} 
\label{fig3}
\end{figure}

We plot in Fig.~\ref{fig3} the spatial profiles of the atomic probability density $|\psi(\mathbf{r},t)|^2$ in the fiber transverse plane for different evolution times. Initially, at $t=0$, the wave packet is well localized, see Fig.~\ref{fig3}(a). We observe from Fig.~\ref{fig3}(b) that for $t\sim 20$~$\mu$s, the wave packet has already spread significantly and $|\psi(\mathbf{r},t)|^2$ is delocalized along a circle in the fiber transverse plane. Figures \ref{fig3}(c) and \ref{fig3}(d) show that, when the evolution time $t$ is about 100~$\mu$s or 200~$\mu$s, the wave packet partially relocalizes in the form of 4 and 2 sub-packets, respectively. These regions are the regions of $T_{\rm rev}/8$ and $T_{\rm rev}/4$ fractional revivals \cite{Averbukh,revival}. According to Figs. \ref{fig3}(e) and \ref{fig3}(f), the probability density $|\psi(\mathbf{r},t)|^2$ reforms into a structure with a single dominant peak when the evolution time $t$ is about 400~$\mu$s or 800~$\mu$s. The structure in Fig.~\ref{fig3}(e) is the result of the half revival realized at $T_{\mathrm{rev}}/2$. Near this time, the wave packet reforms with the original periodicity, but the phase of its orbiting motion differs from the initial wave packet \cite{revival}. The structure in Fig.~\ref{fig3}(f) is the result of the full revival at $T_{\mathrm{rev}}$. Near this time, the wave packet reforms with the original periodicity. Its peak value is reduced and its spread is increased with respect to the original wave packet due to cubic and higher order terms in the dispersion relation. In the special case where $T_{\mathrm{rev}}/T_{\mathrm{rot}}$ is an integer, the full revival is in phase with the initial time development \cite{revival}. 

In order to experimentally reveal the predicted wave packet dynamics, we propose to probe the orbiting atom by a weak, resonant, quasi-linearly polarized guided field $\mathbf{E}_p(\mathbf{r})$ of frequency $\omega_p$. Neglecting the effect of the fiber on the spontaneous emission rate and the detuning~\cite{LeKien05c}, the rate of the scattering from the atom is proportional to the overlap between the atomic wave packet and the probe field,
\begin{equation}
\gamma_{\mathrm{sca}}(t)\propto\int  |\psi(\mathbf{r},t)|^2 |\mathbf{E}_p(\mathbf{r})|^2d\mathbf{r}~.
\label{10}
\end{equation}
Assume that the main axis of the polarization of the probe guided field is aligned at the azimuthal angle $\varphi=0$. 
For the following, we take advantage of the fact that the intensity of the quasi-linearly polarized fundamental-mode guided probe field exhibits an azimuthal modulation according to  \cite{fibermode}
\begin{align}
|\mathbf{E}_p(\mathbf{r})|^2\propto~ & |e_\varphi^{(\omega_p)}(r)|^2+\left[|e_r^{(\omega_p)}(r)|^2-|e_\varphi^{(\omega_p)}(r)|^2\right. \nonumber \\ & \left. +|e_z^{(\omega_p)}(r)|^2 \right]\cos^2 \varphi~,
\label{11}
\end{align}
where $e_r^{(\omega_p)}$, $e_\varphi^{(\omega_p)}$, and $e_z^{(\omega_p)}$ are the cylindrical components of the mode profile vector function ${\bf e}^{(\omega_p)}$ \cite{fiber books, fibermode}.
Hence, we find
\begin{equation}
\gamma_{\mathrm{sca}}(t)\propto  
B+\sum_{m} c_{m-1} c_{m+1} V_m \cos\left(2\Delta\mathcal{E}_mt/\hbar\right),
\label{13}
\end{equation}
where 
$
\Delta\mathcal{E}_m=(\mathcal{E}_{m+1}-\mathcal{E}_{m-1})/2
$ and the coefficients $B$ and $V_m$ are
\begin{eqnarray}
B&=&\sum_{m} c_m^2\int\limits_a^\infty  u_m^2(|e_r^{(\omega_p)}|^2+|e_\varphi^{(\omega_p)}|^2+|e_z^{(\omega_p)}|^2)\, dr,
\nonumber\\
V_m&=&\int\limits_a^\infty u_{m-1} u_{m+1}(|e_r^{(\omega_p)}|^2-|e_\varphi^{(\omega_p)}|^2+|e_z^{(\omega_p)}|^2)\, dr.\quad
\label{14}
\end{eqnarray}

\begin{figure}
\begin{center}
  \includegraphics{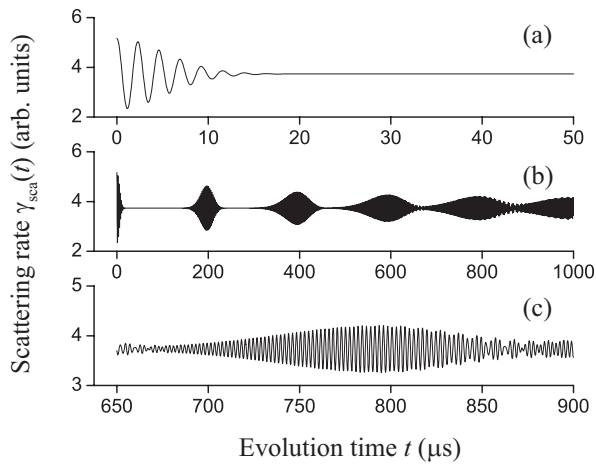}
 \end{center}
\caption{Time dependence of the scattering rate  
$\gamma_{\mathrm{sca}}(t)$ in the ranges from 0 to 50 $\mu$s (a),
from 0 to 1000 $\mu$s (b), and from 650 to 900 $\mu$s (c).
The polarization of the guided probe field is aligned at the azimuthal angle $\varphi=0$.
} 
\label{fig4}
\end{figure}

We plot in Fig.~\ref{fig4} the time dependence of  $\gamma_{\mathrm{sca}}(t)$. Figure \ref{fig4}(a) shows that $\gamma_{\mathrm{sca}}(t)$ oscillates with an initial visibility of almost 40\%. The oscillations result from the classical rotation of the atomic wave packet around the nanofiber. With time, the modulation amplitude reduces and $\gamma_{\mathrm{sca}}(t)$ reaches a quasi-stationary value for $t\sim 15~\mu$s. The period of oscillations of $\gamma_{\mathrm{sca}}(t)$ as obtained from the numerical evaluation is $T_{\mathrm{osc}}^{(\mathrm{sca})}\cong 2.33$ $\mu$s, one half of the classical atomic rotation period $T_{\mathrm{rot}} \cong 4.67$ $\mu$s. This reduction is due to the fact that the intensity of the probe field is symmetric with respect to the reflection  $\{r,\varphi\} \to \{r,\varphi+\pi\}$ in the fiber transverse plane, see Eq.~(\ref{11}). In Fig.~\ref{fig4}(b), clear-cut resumptions of the oscillation of the scattering rate $\gamma_{\mathrm{sca}}(t)$ appear when the evolution time is about 200~$\mu$s, 400~$\mu$s, 600~$\mu$s, and 800~$\mu$s, corresponding to $T_{\mathrm{rev}}/4$, $T_{\mathrm{rev}}/2$,  $3T_{\mathrm{rev}}/4$, and  $T_{\mathrm{rev}}$, respectively. As a specific characteristic of our probing scheme, the fractional revival of the wave packet at $T_{\rm rev}/8$, see Fig.~\ref{fig3}(c), and the fractional revivals at odd multiples of $T_{\rm rev}/8$ as well as all higher order fractional revivals \cite{revival} do not give rise to a modulation of $\gamma_{\mathrm{sca}}(t)$. This can be readily understood considering the four-fold azimuthal symmetry of $|\psi(\mathbf{r},t)|^2$ for these fractional revivals. In conjunction with the $\cos^2\varphi$-dependence of the probe field intensity, one can easily show that the modulation amplitude of $\gamma_{\mathrm{sca}}(t)$ in Eq.~(\ref{10}) vanishes. Figure \ref{fig4}(c) shows a zoom of the signal around $T_{\mathrm{rev}}$. The visibility  of the scattering signal is reduced to about $1/3$ of the initial visibility, revealing the increased spread of the wave packet upon revival, see Fig.~\ref{fig3}(f).

In order to further link the wave packet dynamics with the predicted signal, $\gamma_{\mathrm{sca}}(t)$, we expand $\Delta \mathcal{E}_{m}$ in Eq.~(\ref{13}) into a Taylor series of the second order around the central azimuthal quantum number $m_0$ and find 
\begin{equation}
\Delta\mathcal{E}_m\cong \mathcal{E}'_{m_0}+\mathcal{E}''_{m_0}(m-m_0)~.
\label{16}
\end{equation}
The first term in Eq.~(\ref{16}) leads to the oscillations of the scattering rate with the period $T_{\mathrm{osc}}^{(\mathrm{sca})}=\pi\hbar/\mathcal{E}'_{m_0}=T_{\mathrm{rot}}/2$. This period is the same for the initial and resumed oscillations of the scattering rate. Note that this also holds for the fractional revivals at $T_{\rm rev}/4$ and $3T_{\rm rev}/4$ where two diametric wave packets orbit around the nanofiber. The second term in Eq.~(\ref{16}) leads to the falloff and the resumptions of oscillations in the scattering rate. We define the falloff time of $\gamma_{\mathrm{sca}}(t)$ as the time at which the spread of phase differences between various oscillatory terms in Eq.~(\ref{13}) is about $\pi$ and find $T_{\mathrm{fall}}^{(\mathrm{sca})}=\pi\hbar/(2|\mathcal{E}''_{m_0}|\Delta m)=\sqrt{\pi}T_{\mathrm{coll}}/4$. For the resumption time of $\gamma_{\mathrm{sca}}(t)$, we obtain $T_{\mathrm{resume}}^{(\mathrm{sca})}=\pi\hbar/|\mathcal{E}''_{m_0}|=T_{\mathrm{rev}}/4$. Based on the parameters used in our numerical calculations, we find $T_{\mathrm{osc}}^{(\mathrm{sca})}\cong 2.33$ $\mu$s, $T_{\mathrm{fall}}^{(\mathrm{sca})}\cong 16.5$ $\mu$s, and $T_{\mathrm{resume}}^{(\mathrm{sca})}\cong 199$~$\mu$s, in agreement with Fig.~\ref{fig4}.

The observation of the predicted quantum dynamical effects for cold atoms orbiting in a light-induced potential surrounding an optical nanofiber is within the scope of current nanofiber-based quantum optics experiments. In particular, loading of atoms and wave packet preparation in stable orbits around the nanofiber should be possible by starting with a stationary trap in which the repulsion of the atoms from the nanofiber surface is accomplished by means of a blue-detuned nanofiber-guided light field \cite{Dowling,twocolors,twocolor experiment,Goban12}. After laser-cooling of the atoms into the vibrational ground state of the two-color optical potential, the latter can then be set into rotation using polarization modulators while suitably turning off the blue-detuned field, eventually preparing the trap and wave packet proposed here. This system can be seen as a two-dimensional ``artificial atom'' where the orbiting atom and the nanofiber take the role of the valence electron and the ion core, respectively, while the Coulomb attraction is replaced by a tunable light-induced potential. In view of the tunability of the dispersion relation, this would then implement a versatile experimental platform for the investigation of quantum wave packet dynamics.

We thank J.~Burgd\"orfer, J.~I.~Cirac, and J.~Dalibard for helpful comments and discussions. Financial support by the ESF (EURYI Award) and the Wolfgang Pauli Institute is gratefully acknowledged.


\begin{thebibliography}{99}

\bibitem{Schroedinger26} E. Schr\"odinger, Naturwissenschaften \textbf{14}, 664 (1926).

\bibitem{Stroud1986} J. Parker and C.~R. Stroud, Jr., Phys. Rev. Lett. \textbf{56}, 716 (1986); Phys. Scr. T \textbf{12}, 70 (1986).

\bibitem{revival} R.~W. Robinett, Phys. Rep. \textbf{392}, 1 (2004).

\bibitem{Will10} S. Will, T. Best, U. Schneider, L. Hackerm\"uller, D.-S. L\"uhmann, and I. Bloch, Nature \textbf{465}, 197 (2010).

\bibitem{Wyker12} B. Wyker, S. Ye, F.~B. Dunning, S. Yoshida, C.~O. Reinhold, and J. Burgd\"orfer, Phys. Rev. Lett. \textbf{108}, 043001 (2012).

\bibitem{Keitel} O.~D. Skoromnik, I.~D. Feranchuk, C.~H. Keitel, arXiv:1209.1939 (2012).

\bibitem{Dowling} J.~P. Dowling, and J. Gea-Banacloche, Adv. At. Mol. Opt. Phys. \textbf{37}, 1 (1996).

\bibitem{onecolor} V.~I. Balykin, K. Hakuta, Fam Le Kien, J.~Q. Liang, and M. Morinaga, Phys. Rev. A \textbf{70}, 011401(R) (2004). 

\bibitem{Averbukh} I.~Sh. Averbukh and N.~F. Perelman,  Phys. Lett. \textbf{139}, 449 (1989).

\bibitem{fiber books} A.~W. Snyder and J.~D. Love, \textit{Optical Waveguide Theory} (Chapman and Hall, New York, 1983).

\bibitem{twocolors} Fam Le Kien, V.~I. Balykin, and K. Hakuta, Phys. Rev. A \textbf{70}, 063403 (2004).

\bibitem{twocolor experiment} E. Vetsch, D. Reitz, G. Sagu\'e, R. Schmidt, S.~T. Dawkins, and A. Rauschenbeutel, Phys. Rev. Lett. \textbf{104}, 203603 (2010).

\bibitem{Goban12} A. Goban \textit{et al.}, Phys. Rev. Lett. \textbf{109}, 033603 (2012).

\bibitem{LeKien05c} Fam Le Kien, S. Dutta Gupta, V.~I. Balykin, and K. Hakuta, Phys. Rev. A \textbf{72}, 032509 (2005).

\bibitem{fibermode} Fam Le Kien, J.~Q. Liang, K. Hakuta, and V.~I. Balykin, Opt. Commun. \textbf{242}, 445 (2004).





\end{thebibliography}
\end{document}